# SMEs' Confidentiality Concerns for Security Information Sharing

Alireza Shojaifar[1,2], Samuel A. Fricker[1,3]

[1] FHNW, IIT, 5210 Windisch, Switzerland
(alireza.shojaifar|samuel.fricker)@fhnw.ch
[2] Utrecht University, Dept. of Information and Computing Sciences, Utrecht, Netherlands
a.shojaifar@uu.nl
[3] Blekinge Institute of Technology, SERL-Sweden, 371 79 Karlskrona, Sweden
samuel.fricker@bth.se

**Abstract.** Small and medium-sized enterprises (SME) are considered an essential part of the EU economy; however, highly vulnerable to cyber-attacks. SMEs have specific characteristics which separate them from large companies and influence their adoption of good cybersecurity practices. To mitigate the SMEs' cybersecurity adoption issues and raise their awareness of cyber threats, we have designed a self-paced security assessment and capability improvement method, CYSEC. CYSEC is a security awareness and training method that utilises self-reporting questionnaires to collect companies' information about cybersecurity awareness, practices, and vulnerabilities to generate automated recommendations for counselling. However, confidentiality concerns about cybersecurity information have an impact on companies' willingness to share their information. Security information sharing decreases the risk of incidents and increases users' self-efficacy in security awareness programs. This paper presents the results of semi-structured interviews with seven chief information security officers (CISOs) of SMEs to evaluate the impact of online consent communication on motivation for information sharing. The results were analysed in respect of the Self-Determination Theory (SDT). The findings demonstrate that online consent with multiple options for indicating a suitable level of agreement improved motivation for information sharing. This allows many SMEs to participate in security information sharing activities and supports security experts to have a better overview of common vulnerabilities.

## 1     Introduction

Small and medium-sized enterprises (SMEs) have a considerable diversity and form the backbone of the EU's economy [12]. However, although they need to deal with a similar level of cybersecurity risk as large companies, information security is not always a priority [17,24]. Moreover, the lack of written security policy, financial resources, and security expertise are other operational constraints and make SMEs more vulnerable [11,13]. To having an effective understanding of security policy and fostering a security culture, providing appropriate training and awareness tools specifically for small enterprises is necessary [11].

Training and awareness programs are the most commonly suggested approaches in the literature for security policy compliance, and they can alleviate employees' limited knowledge of cybersecurity [20]. Systematic training programs are a good means of facilitating continuous information security communication in organisations. Information security training should apply content and approaches that enable and motivate learners to systematic cognitive processing of information they receive in training [20].

CYSEC is a self-paced SME-specific training and awareness method that provides training by automating the elements of security communication between employees and a security expert [2]. Since there is a resistance to changing cybersecurity behaviour and adopting security tools by employees [16], the method implements the motivational constructs—namely, the needs for autonomy, competence, and relatedness—in the self-determination theory (SDT) [4,29] to motivate learners to adopt advice. CYSEC provides training, recommendations, and relevant hands-on skills based on SMEs' answers to the self-assessment questions to enable them to become more self-determined.

Security information sharing is a challenge for companies, and they are reluctant to share their information and report their incidents [9,15,30]. Fear of negative publicity and competitive disadvantage, believing that the chance of a successful prosecution is not high, believing that the cyber incident was not severe enough to be reported, and a lack of motivation and trust are some of the major hindrance to information sharing activities [15,30]. However, security information sharing is a significant measure to reduce the risks of similar incidents and develop a better understanding of the risks facing a community [21,30]. The European Network and Information Security Agency (ENISA) [21] explains that the nature of cyber incidents and attacks is borderless. To support the management of threats and vulnerabilities in the community of cybersecurity, the exchange of data and cross-border cooperation is necessary [21]. ENISA indicates that trust is the critical element to enhance security information sharing. Therefore, the actual usage of CYSEC requires further investigation.

The current study aims to evaluate the impact of online consent communicating on SME's chief information security officers (CISOs) motivation for security information sharing. Taking approaches that motivate users to adopt security recommendations can support the effectiveness of cybersecurity communication [1]. The semi-structured interview method was selected for data collection about behavioural motivation. The data was qualitatively analysed based on a proposed theoretical model by Yoon and Rolland [26] for explaining knowledge-sharing behaviours in virtual communities. The model studies the effect of basic psychological needs in SDT (autonomy, competence, and relatedness) and two antecedents of the basic needs: familiarity and anonymity. Familiarity refers to an individual's understanding of the environment and increases the trust of other people [8]. Anonymity refers to the inability of others to identify an individual or for others to identify one's self [6] and may influence individuals' knowledge sharing behaviour in a virtual community [26].

Our study results indicate that SDT and two antecedents (familiarity and anonymity) account for motivation for security information-sharing behaviour, and online consent has a positive impact on CISOs' motivation. The online consent increased users' trust and provided value for SMEs to make choices and decisions about the suitable level of relatedness for sharing information.

The remainder of the paper is organised as follows. Section 2 presents the research background, the theoretical model, and the research prototype. Section 3 describes the design of our study. Section 4 presents the analysis approach and the answer to the research question. Section 5 discusses the significance of the results and the threats to validity. Section 6 summarises and concludes.

## 2     Research Background

Security information sharing has been identified to increase end-users' self-efficacy in security awareness programs [9, 19]. Security information sharing means *the exchange of network and information security-related information such as risks, vulnerabilities, threats, internal security issues, and good practice* [21]. Security information should be shared to understand the risks facing the community and any related significant information infrastructure and reduce the risk of incidents.

Confidentiality concerns and the lack of incentives prevent companies to share security information [9,30]. Geer et al. [30] state *individual companies might have some rudimentary understanding of their own information security health, but we have no aggregate basis for public policy because organisations do not share their data*. The companies' confidentiality concerns include worries about reputation, losing customers, fears of misuse of the information, and strong emotional relatedness to the organisational data. These concerns exist even if security information is anonymised.

Trust influences a user's willingness to share knowledge [26,28] and security information [21]. Hosmer [25] defines trust as *the expectation by one person, group, or firm of ethically justifiable behaviour on the part of the other person, group, or firm in a joint endeavour or economic exchange.* Some arrangements could mitigate confidentiality concerns relevant to trust issues. The arrangements include to (1) give control of information to the company which shared it, (2) agree about how to use and protect shared information, (3) preserve data anonymity, and (4) develop standard terms for communicating information [9]. Deci et al. [5] explain that also autonomy support, including the offering of choice and relevant information, impacts trust.

The Self-Determination Theory (SDT) has been proposed as a theoretical framework to study humans' motivational dynamics and consequent behaviours [4,5,29]. People have different levels and orientations of motivation. Self-determination in SDT is defined as: *the capacity to choose and to have those choices, rather than reinforcement contingencies, drives, or any other forces or pressures, to be the determinants of one's actions. But self-determination is more than capacity. It is also a need.* Deci and Ryan [7] have hypothesised a basic, innate aptitude to be self-determining that leads humans and organisations to engage in desirable behaviours.

SDT assumes that the satisfaction of humans' basic psychological needs - autonomy, competence, and relatedness – leads to self-motivation and positive outcomes [27]. Autonomy refers to a desire to engage in activities with a choice of freedom. Competence implies that individuals have a desire to interact effectively with the environment for producing desired outcomes and preventing undesired events. Relatedness reflects a sense of belongingness and connectedness to others or a social environment.

To explain knowledge-sharing behaviours, Yoon and Rolland [26] extended SDT with two antecedents, familiarity and anonymity. Familiarity refers to an individual's understanding of an environment based on the prior experience and learning of the what, who, how, and when of what is happening [8]. Familiarity may improve perceived competence, the feeling of relatedness [26]. Anonymity refers to the inability of others to identify a person or for others to identify one's self [6]. It can reduce social barriers and allow group members to contribute their opinions [26]. Anonymity may impact on autonomy and the feeling of relatedness. Fig. 1 shows the complete model.

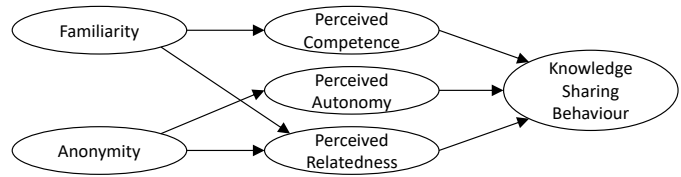

Fig. 1. Research model (Yoon and Rolland [26])

We applied Yoon and Rolland [26] model to evaluate the impact of the motivational factors in online consent on the security information-sharing activities of SME CISOs. We asked SMEs to use self-assessment questionnaires to collect security information and share with a community of security experts and other SMEs. The consent provides choices and the opportunity for CISOs to exert control over information sharing. Through the choices, CISOs can define their relatedness to the tool and the community. Each choice gives information and explains how and where the shared information will be used to increase users' familiarity with the data usage environment. The consent emphasises that the shared information will be used anonymously.

The consent form included three choices based on three levels of relatedness and agreement. 1) disagreement to share security information. 2) agreement to automated processing of security information for recommendations of cybersecurity improvements in their company. 3) agreement to share security information for cybersecurity research. The form includes the choice of anonymity. It supports familiarity by elaborating on the usage of security information to reduce the complexity for new users and enhance the users' competence. Fig. 2 shows the consent form

> **Replying to questionnaires is a part of your self-evaluation approach.**
> **However, you can improve adherence to good cybersecurity practices in the CYSEC community.**
> ○ I use the tool for testing purposes or my personal training (my answers may be not accurate).
>   *By choosing this item, you are using the tool only for the training and awareness purposes, such as reading the training and awareness content and using the embedded training links. And your answers will be removed from the tool.*
> ○ I use the tool to improve the cybersecurity in my SME with the automated feedback and recommendations it generates (I confirm that my answers are correct).
>   *By choosing this item, the stored answers are used to generate recommendations, feedback, and KPIs.*
>
> ☐ My anonymised answers may be used in the CYSEC community for further research.
>   *By choosing this item, the stored data is used: (1) for your SME cybersecurity improvement and (2) anonymously for conducting research in **FHNW University** for the CYSEC community to improve the cybersecurity coping mechanisms, for instance, generating recommendations based on all CYSEC partners' capability to make a backup.*
>
>                              [ Agree ]

Fig. 2. Screenshot of the Online Consent prototype

## 3  Method

This study aims at finding out the impact of online consent communicating on SME's CISOs motivation for sensitive information sharing behaviour. Semi-structured interviews [10] were applied to conduct the empirical part of this study. The interview is one of the most frequently used methods and the most significant sources of data in empirical studies in software engineering. Interviews provide researchers with important insights into the quality and usability of artefacts since much of the knowledge that is of interest is only available in the minds of users [14]. The same method (interview with CISOs, and key informants who are Network and Information Security experts) has already been used in the context of cybersecurity [22] and security information sharing [9].

A theoretical model based on SDT (Fig. 1) was chosen to analyse the CISOs' information-sharing motivation and whether online consent can motivate them. When we have a developed theory as a template, the model of generalisation is the analytic generalisation [10]. The study seeks answers to this research question: *Do the choice of anonymity and the elaboration of how shared information will be used motivate CISOs of SMEs to share security information?* Security information sharing is a necessary measure in the context of cybersecurity [18,21] and for SMEs [3,23]. We are studying how motivational constructs, controlling over data through choices, online agreement, and familiarity with the usage of shared information, can impact information-sharing behaviour. Recorded interviews were analysed based on content analysis (interviewees' argumentations) and theoretical cause-effect relationships [10].

At first, a pilot study, including three interviews with three SMEs' CISOs (project partners), has been conducted—the pilot study allowed to identify and resolve initial problems in the interview questions and the online consent design. The selection of the subjects was based on the availability of the SMEs. There were twelve SMEs (four project partners and eight open call partners), and seven of them participated in the interviews. The participating SMEs came from five EU countries, and all were active in the IT industry. All of them implemented some security controls, including password management, basic approaches for privacy protection, using firewalls, two-factor authentication, cloud security features, and anti-virus installation. One person from each SME was interviewed. The people interviewed were chosen because they all were CISOs or senior managers, and all have been involved in cybersecurity tasks within their companies. All were college graduates and had several years of experience in security. One of the interviewed people was a cybersecurity expert and provided a more in-depth perspective on the importance of security information sharing, the necessity for an agreement, and anonymity. Table 1 presents the SMEs' demographics. In the European Union, companies are considered to be SMEs if they have fewer than 250 employees and an annual turnover of less than € 50 million [31].

Interviews were conducted face-to-face when possible. For four SMEs, a request for the online interview has been sent. All interviewees had the possibility to find a suitable time. In the online interviews, the screen of the interviewer's computer was shared, and the interviewees were able to see and read the content and had enough time to think about the answers. All the interviews were conducted without distraction. Each interview started with an explanation of the study. Then we presented each interviewee with a screenshot of the online consent and asked them about their understanding of it. All interviewees understood the idea and content. To collecting honest responses, the researcher emphasised that the collected data would be applied anonymously for academic purposes and then obtained the subjects' consent. In the end, a summary of the key findings and answers presented to the interviewees. All interviews were recorded and transcribed.

**Table 1**. SMEs' Demographics

| ID | Org. size | Offices | Maturity | Structure |
|---|---|---|---|---|
| 1 | Small | 2 | Some controls implemented | CEO, security team, employees |
| 2 | Small[1] | 1 | Some controls implemented | Professors, manager, security team, users |
| 3 | Medium | 3 | Some controls implemented | CEO, security team, employees |
| 4 | Small | 2 | Some controls implemented | CEO, security manager, employees, behavioural scientist |
| 5 | Small | 2 | Some controls implemented | CEO, employees |
| 6 | Small | 1 | Some controls implemented | CEO, employees |
| 7 | Small | 2 | Some controls implemented | CEO, security team, employees |

## 4    Analysis of the Interview Results

For answering the research question, we studied the impact of anonymity choice and elaboration of data use on SME CISOs' motivation for security information sharing. The interview transcripts showed that both design elements of the online consent form affected the CISOs' motivation for security information sharing. They supported relatedness, autonomy, and competence, and enhanced the CISOs' trust perception. The study participants were motivated to share security information when they perceived that they had control of the communication, and the information was securely stored.

**Security information sharing behaviour.** Through the interviews, it became clear that the agreement form encouraged the CISOs' information sharing with the tool. ID7 emphasised that the agreement was not only useful but also legally necessary. ID3 and ID7 stated that the agreement positively affected their trust. ID3: *"it has a positive effect on trust because it shows that you take care of the data process and make it clear."* ID7: *"the online consent impacts on trust and shows me that there are thoughts, conditions, and efforts to provide different options and approaches for disclosure."* ID1: *"with this agreement, I feel safer."*

**Role of autonomy through choice**. The analysis of the study participants' arguments showed that the autonomy offered by the choice of sharing security information influenced their information-sharing intention. ID7: *"providing options show me that these people know what they are asking for and give you options."* All interviewees recognised the importance of security information sharing for receiving better advice;

---

[1] The organisation 2 was a university-hosted start-up.

however, some of them selected the third choice (sharing information for research). ID1: *"I may change my answer later."* ID7: *"I need to check with my boss. Do I still have the ability to edit my selection? I can decide and let you use the answers, but until the end of the duration of the project."* The respondents were asked whether they wanted to add a new option to the online consent, but no new option was suggested.

**Role of familiarity through elaboration of security information use.** Improving the CISOs' familiarity with how security information would be used positively affected competence and relatedness. All except ID6 wanted to have a clear description of how their information would be used. ID6 stated: *"the text is clear and understandable, and it improves my awareness."* IDs 1, 3, 4, 7 emphasised that the agreement has not provided sufficient information. When asked why, ID1 stated: *"I do not know how my company information is stored; also, it should be stated if we can change our answer later."* ID3: *"I assumed that after generating recommendations, the data should be destroyed."* ID7: *"it should be clearly stated if the data will be used in the future and after the project."* ID4 also wanted to know more about the security information recipient "FHNW" indicated on the consent form. The interviewees were also asked to state if they wanted to know how the information is processed. Most interviewees stated that their organisations only wanted to have the results of the analysis, i.e. the tool's recommendations. ID2: *"I do not care how and when my data will be used for the research. I just want to have the results."* ID5 and ID6 wanted to know, however, how their information would be processed.

The content of the agreement was perceived to be confusing for some of the subjects. Some of the interviewees suggested modifications to the content. ID2: *"the second option should be rephrased: [try to answer the questionnaires to the best of your knowledge to help us give you more accurate recommendations]."* *"Options should be re-arranged: Options #2 and #3 should be separated from option #1."* *"Option #3 should be rephrased, something like [if you allow us to collect your answer, we will be able to improve the tool, provide you better analysis, and better help you in the future. (Yes or No)]?"* ID4: *"rephrasing can clarify the message because I do not know if I select option #1, I will receive a recommendation."* ID7: *"I think giving an option to SMEs that indicates my answer may or may not be accurate can demote the whole."*

**Role of anonymity through security information anonymisation**. Anonymising the security information could influence perceived relatedness and autonomy and, in turn, encourage security information sharing. The analysis of the interviewees' arguments showed that all believed that anonymity would reduce the risks of sharing information. They felt more secure when the tool support anonymity. ID2: *"if my data is anonymised, I don't care how my data will be used."* ID1: *"anonymised data sharing shows that it is safe."* ID7: *"I presume that when you put stress on anonymity in the third option, it can imply that the second option is not anonymised. I assume that even for other usages (KPI, recommendations), SMEs should not be recognisable."*

The interviewees would not share security information that would expose details about their organisation, hence would break their anonymity. ID3: *"consent cannot change my opinion; I am not answering the textbox questions."* ID7: *"I know that Yes/No or multi-choice questions can be used for the statistical analysis; however, any question that refers more to deterministic answers, I don't want to answer."*

## 5 Discussion

### 5.1 Security information sharing

Security information sharing is widely acknowledged [9,21,30]; however, confidentiality worries, lack of incentives, and lack of trust lead the companies to avoid sharing information and reporting vulnerabilities [9,30]. To motivating companies to share their security information, attention to arrangements such as giving control of information to the company which shared information, having an agreement, and preserving data anonymity is necessary [9].

In this study, based on a theoretical model for knowledge-sharing behaviour in virtual communities [26], we have evaluated the impact of online consent communicating on SME CISOs' motivation for information sharing. The model [26] extended the self-determination theory and included two antecedents (familiarity and anonymity) on basic psychological needs (perceived autonomy, perceived competence, and perceived relatedness). Yoon and Rolland [26] study indicates that perceived autonomy does not influence knowledge-sharing behaviours in virtual communities since a virtual community is a voluntary environment that is not controlled by anyone else. However, our findings show that in the context of security information sharing, users' perception of controlling over information sharing increased their motivation, and providing choices enabled users to have selective permission controls. This finding is consistent with the previous study [9]. Moreover, Yoon and Rolland [26] study shows that anonymity has a negative impact on knowledge-sharing activities since the anonymity in a virtual community can be used to attack the opinions of other people, and *in a highly anonymous environment, individuals may think about other people's reactions to their opinion*. In our study, users emphasised that preserving anonymity is essential. Although our study is based only on qualitative findings and a small sample, we can explain the anonymity based on the perception of altruism [28] and the risk of information misuse [3].

In CYSEC, the self-assessment questionnaires are used to collect security information (including cybersecurity awareness, practices, and vulnerabilities) and share with a community of security experts and other SMEs. The results demonstrated that online consent with the choice of anonymity and the elaboration of how shared information is used motivated CISOs of the SMEs to share their information. Also, we discovered that CISOs would not share security information that would expose details. For future research, the other legal and economic incentives [18] should be considered, and not only CISOs opinions but also employees' viewpoints should be studied.

### 5.2 Study Limitation

This study has some limitations. One criterion influencing the sufficiency of the interviews was saturation. The saturation point is reached when no new information is gathered, or the subjects' viewpoints are repeated [14]. Due to the small sample size, we could not reliably validate saturation and implement a statistical analysis in our study. The study is based on seven interviewed persons from seven SMEs that were active in the IT industry, which limits generalizability. Further research with a larger sample and a diversity of SMEs could reveal more robust results and provide more insights into the influence of the industry type on the SME engagement in security information sharing. Second, since the study is based on the CISOs and senior managers' viewpoints of security information sharing, our study lacks the view of SMEs' employees. To having a wider perspective, the views of SMEs' employees are needed.

## 6      Summary Conclusions

The paper has evaluated the impact of online consent communicating on motivating CISOs of SMEs for security information sharing. This study followed a deductive approach and tested constructs drawn on the Self-Determination Theory (SDT) as well as two antecedents of SDT constructs (familiarity and anonymity) to evaluate the impact of the online consent on the security information sharing motivation. We applied semi-structured interviews with seven CISOs from seven SMEs for data collection. The study results indicate that online consent increased CISOs' trust and had a positive impact on security information sharing intention. The consent supports familiarity with the environment through the elaboration of security information usage. Moreover, online consent considers the role of anonymity and autonomy through security information anonymisation and the choice of sharing information.

## Acknowledgments

This work has received funding from the European Union's Horizon 2020 research and innovation programme under grant agreements No. 740787 (SMESEC), No. 883588 (GEIGER), and the Swiss State Secretariat for Education, Research and Innovation (SERI) under contract number 17.00067. The opinions expressed and arguments employed herein do not necessarily reflect the official views of these funding bodies.